\documentclass[runningheads,fleqn]{svmult}
\usepackage{makeidx} 
\usepackage{graphicx}
\usepackage{subeqnar}
\usepackage{multicol}
\usepackage{taphys}  
\makeindex           
\usepackage{amstex}  
\mathindent\parindent 
\newcommand{\ud}{\text{d}}
\newcommand{\ui}{\text{i}}
\newcommand{\ue}{\text{e}}
\newcommand{\vecal}{\boldsymbol{\alpha}} 
\newcommand{\vecsig}{\boldsymbol{\sigma}}
\newcommand{\vecxi}{\boldsymbol{\xi}}
\begin{document}
\title*{The Gutzwiller Trace Formula for\\ 
Quantum Systems with Spin}
\toctitle{The Gutzwiller Trace Formula for
\protect\newline Quantum Systems with Spin}
\titlerunning{Trace Formula with Spin}
\author{Jens Bolte}
\institute{Abteilung Theoretische Physik, Universit\"at Ulm,
D-89069 Ulm, Germany}

\maketitle           

\begin{abstract}
The Gutzwiller trace formula provides a semiclassical approximation
for the density of states of a quantum system in terms of classical
periodic orbits. In its original form Gutzwiller derived the trace
formula for quantum systems without spin. We will discuss the 
modifications that arise for quantum systems with both translational
and spin degrees of freedom and which are either described by Pauli-
or Dirac-Hamiltonians. In addition, spectral densities weighted by
expectation values of observables will be considered. It turns out 
that in all cases the semiclassical approximation yields sums over
periodic orbits of the translational motion. Spin contributes via
weight factors that take a spin precession along the translational
orbits into account.
\end{abstract}

Thirty years ago, after several intermediate steps the Gutzwiller trace 
formula \cite{Gut71,Gut90} resulted from a detailed semiclassical 
investigation of the time evolution in quantum mechanics. It opened the way
for the application of semiclassical methods to many problems that were
so far believed to lie beyond the capability of semiclassics. The most
prominent example being semiclassical quantisation rules for classically
non-integrable systems, for which Einstein \cite{Ein17} already in 1917 had 
shown that the usual Bohr-Sommerfeld type quantisation methods fail.
Gutzwiller, however, devised a semiclassical expansion for the density of 
states in terms of a sum over the classical periodic orbits for a huge class
of quantum systems, which in particular includes classically chaotic systems.
Shortly afterwards, but seemingly independently, also mathematicians became
interested in such trace formulae. They devised mathematical proofs for various
versions of the trace formula, beginning with the work of Colin de V\`erdiere
\cite{CdV73}, and Duistermaat and Guillemin \cite{DuiGui75}.

What was lacking so far, however, was a trace formula for quantum systems
with a priori non-classical degrees of freedom as, e.g., spin. In such
cases it is not immediately clear what the corresponding classical system is
whose periodic orbits enter the trace formula, and how the non-classical 
degrees of freedom have to be taken into account. Even for systems with a 
classically integrable translational part there do not exist Bohr-Sommerfeld 
(or EBK) type quantisation rules for the eigenvalues of a Dirac-Hamiltonian, 
although already in 1932 Pauli \cite{Pau32} began to generalise the WKB method
to the Dirac equation. Pauli's undertaking was only completed in 1963 by 
Rubinow and Keller \cite{RubKel63}, and it took again some 30 years before 
Emmrich and Weinstein \cite{EmmWei96} proved that due to geometric 
obstructions for the Dirac equation EBK-quantisation rules are generally 
impossible. It therefore seems to be especially desirable to have a Gutzwiller
trace formula for Dirac-Hamiltonians available. Below we explain how one 
proceeds, if the Hamiltonian at hand describes a quantum system with a 
spin 1/2 coupled to the translational motion, be it a relativistic or a 
non-relativistic situation. The method that is used was developed in 
\cite{BolKep99}, where most of the details can be found that cannot be given 
here.

\section{Dirac- and Pauli-Hamiltonians}
In the following we will consider relativistic and non-relativistic particles
with mass $m$, charge $e$ and spin 1/2 in external static electromagnetic
fields $\vec{E}(\vec{x}) = -\nabla\varphi(\vec{x})$ and $\vec{B}(\vec{x}) = 
\nabla\times\vec{A}(\vec{x})$. In the relativistic case the quantum dynamics 
are generated by a Dirac-Hamiltonian
\begin{equation}
\label{HD}
\hat H_{\rm D} = c\vecal\cdot \left( \frac{\hbar}{\ui}\nabla - \frac{e}{c}
\vec{A}(\vec{x}) \right) + mc^2\beta + e\varphi(\vec{x})\ ,
\end{equation}
where $\vecal=(\alpha_1,\alpha_2,\alpha_3)$ and $\beta=\alpha_0$ are hermitian
$4\times 4$ matrices satisfying the algebraic relations $\alpha_\mu\alpha_\nu
+\alpha_\nu\alpha_\mu=2\delta_{\mu\nu}$. In Dirac representation they read
\begin{equation}
\label{alpha}
\vecal = \left( \begin{array}{cc} 0&\vecsig \\ \vecsig&0 \end{array}\right) 
\qquad\text{and}\qquad \alpha_0 = \beta = 
\left(\begin{array}{cc} {\bf 1}_2&0 \\ 0&-{\bf 1}_2 \end{array}\right) \ ,
\end{equation}
where $\vecsig=(\sigma_1,\sigma_2,\sigma_3)$ denotes the Pauli matrices
and ${\bf 1}_2$ is a $2\times 2$ unit matrix. As it is well known, in leading
non-relativistic order $\hat H_{\rm D}$ is replaced by the Pauli-Hamiltonian
\begin{equation}
\label{HP}
\hat H_{\rm P} = \left[ \frac{1}{2m} \left( \frac{\hbar}{\ui}\nabla - 
\frac{e}{c} \vec{A}(\vec{x}) \right)^2 + e\varphi(\vec{x}) \right] {\bf 1}_2 - 
\frac{e\hbar}{2mc}\,\vec{B}(\vec{x})\cdot\vecsig \ .
\end{equation}
For further information, see \cite{Tha92}.
We remark that for the following the explicit form of the right-most term
in (\ref{HP}), describing a coupling of spin to the translational degrees
of freedom, is inessential. It can be any operator of the form $\hbar
\vec{C}(\frac{\hbar}{\ui}\nabla,\vec{x})\cdot\vecsig$, where the components of
$\vec{C}(\vec{p},\vec{x})$ are suitable functions on phase space; e.g., 
$\vec{C}(\vec{p},\vec{x}) = \frac{1}{4mc^2 |\vec{x}|}\frac{\ud\varphi
(|\vec{x}|)}{\ud |\vec{x}|}(\vec{x}\times\vec{p})$ would yield a spin-orbit 
coupling in a spherically symmetric potential.

In applications one is sometimes also interested in describing a coupling
of spin and translational degrees of freedom that is semiclassically
strong. In this case one would consider a (Pauli-) Hamiltonian of the form
\begin{equation}
\label{HPalt}
\hat H_{\rm P'} = \left[ \frac{1}{2m} \left( \frac{\hbar}{\ui}\nabla - 
\frac{e}{c}\vec{A}(\vec{x}) \right)^2 + e\varphi(\vec{x}) \right] {\bf 1}_2 + 
\vec{D}\left(\frac{\hbar}{\ui}\nabla,\vec{x}\right)\cdot\vecsig \ ,
\end{equation}
where again the components $D_k(\vec{p},\vec{x})$ are suitable functions 
on phase space (independent of $\hbar$).

\subsection{Weyl Representation}
All of the above Hamiltonians can be represented as Weyl operators, i.e.,
their action on $n$-component spinors $\psi(\vec{x})=(\psi_1(\vec{x}),\dots,
\psi_n(\vec{x}))^T$, where $n=4$ applies to the case of Dirac spinors and 
$n=2$ to Pauli spinors, can be given as
\begin{equation}
\label{HWeyl}
\hat H \psi(\vec{x}) = \frac{1}{(2\pi\hbar)^3}\int\int H\left(\vec{p},
\frac{\vec{x}+\vec{y}}{2}\right)\,\ue^{\frac{\ui}{\hbar}\vec{p}\cdot
(\vec{x}-\vec{y})}\,\psi(\vec{y})\ \ud p\,\ud y \ .
\end{equation}
On the right-hand side the Weyl symbol $H(\vec{p},\vec{x})$ is a function on 
phase space taking values in the hermitian $n\times n$ matrices. For the 
Hamiltonians (\ref{HD}),(\ref{HP}) and (\ref{HPalt}) these arise upon 
replacing $\frac{\hbar}{\ui}\nabla$ by $\vec{p}$.

In the case of the Dirac-Hamiltonian (\ref{HD}) the symbol $H_{\rm D}(\vec{p},
\vec{x})$ is a hermitian $4\times 4$ matrix with the two doubly degenerate
eigenvalues
\begin{equation}
\label{HDeigenv}
H^\pm (\vec{p},\vec{x}) = e\varphi(\vec{x}) \pm \sqrt{(c\vec{p}-e\vec{A}
(\vec{x}))^2 + m^2c^4} \ .
\end{equation}
These eigenvalues can be recognized as the classical relativistic 
Hamiltonians for spinless particles $(+)$ and anti-particles $(-)$. 
The fact that the corresponding eigenspaces are two dimensional reflects 
the quantum mechanical spin 1/2 of the particles and anti-particles. 
More precisely, at each point $(\vec{p},\vec{x})$ of phase space the 
eigenspaces of $H_{\rm D}(\vec{p},\vec{x})$ can be viewed as Hilbert spaces 
of a spin 1/2, one for particles and one for anti-particles. See 
\cite{BolKep99} for further details. 

The situation is similar for the Pauli-Hamiltonian (\ref{HP}), except for
the absence of anti-particles: $H_{\rm P}(\vec{p},\vec{x})$ is a
hermitian $2\times 2$ matrix,
\begin{equation}
\label{HPsymbol}
H_{\rm P}(\vec{p},\vec{x}) = \left[ \frac{1}{2m} \left( \vec{p} - \frac{e}{c}
\vec{A}(\vec{x}) \right)^2 + e\varphi(\vec{x}) \right] {\bf 1}_2 - 
\frac{e\hbar}{2mc}\,\vec{B}(\vec{x})\cdot\vecsig \ ,
\end{equation}
whose classical part (independent of $\hbar$) is proportional to ${\bf 1}_2$.
This principal symbol therefore has one doubly degenerate eigenvalue, 
$H_0(\vec{p},\vec{x})$, which simply is the factor multiplying ${\bf 1}_2$. 
It can readily be identified as the classical non-relativistic Hamiltonian 
for spinless particles. As in the previous case spin is represented by the
corresponding two dimensional eigenspace of the principal symbol, which
in this case is trivial. 

The Hamiltonian (\ref{HPalt}), however, leads to a different interpretation
of its symbol,
\begin{equation}
\label{HPaltsymbol}
H_{\rm P'}(\vec{p},\vec{x}) = \left[ \frac{1}{2m} \left( \vec{p} - \frac{e}{c}
\vec{A}(\vec{x}) \right)^2 + e\varphi(\vec{x}) \right] {\bf 1}_2 + 
\vec{D}(\vec{p},\vec{x})\cdot\vecsig \ ,
\end{equation}
since this is independent of $\hbar$ and thus has to be considered as a
classical quantity in its entirety. Its two eigenvalues
\begin{equation}
\label{HPalteigenv}
H^{\uparrow/\downarrow}(\vec{p},\vec{x}) = \frac{1}{2m} \left( \vec{p} - 
\frac{e}{c}\vec{A}(\vec{x}) \right)^2 + e\varphi(\vec{x}) \pm \left|
\vec{D}(\vec{p},\vec{x})\right|
\end{equation}
are classical non-relativistic Hamiltonians of a particle with fixed spin 
up or down, respectively. Here `up' and `down' are defined with respect to 
the direction of $\vec{D}$ at the respective point $(\vec{p},\vec{x})$ in
phase space. The eigenvalues (\ref{HPalteigenv}) are non-degenerate as 
long as they are different, i.e., away from mode conversion points where 
$\vec{D}(\vec{p},\vec{x}) = 0$. 

From the above discussions one may anticipate that in the semiclassical
considerations of the Hamiltonians (\ref{HD}) and (\ref{HP}) to follow 
(see also \cite{BolKep99}) there will occur classical translational dynamics 
that are uninfluenced by spin, and quantum mechanical spin dynamics driven 
by the classical translational motion. In contrast, in the case of the 
Hamiltonian (\ref{HPalt}) the classical translational motion will depend on 
the (fixed) direction of the spin and there will be no additional spin 
dynamics.

\subsection{Spectra}
Gutzwiller's trace formula provides a semiclassical expansion for the quantum
mechanical density of states. Since this quantity a priori requires a
Hamiltonian with a discrete spectrum, we now want to discuss the spectra
of the Hamiltonians (\ref{HD}), (\ref{HP}) and (\ref{HPalt}). 

Dirac-Hamiltonians typically possess continuous spectra. Indeed, if the
electromagnetic fields vanish as $|\vec{x}|\to\infty$, $\hat H_{\rm D}$ has 
an essential spectrum consisting of two half axes, $(-\infty,-mc^2]\cup 
[+mc^2,+\infty)$, see \cite{Tha92} for precise statements. 
Thus, since we are interested in the discrete spectrum of a Hamiltonian, 
we have to localise in energy to within the gap $(-mc^2,+mc^2)$ of the 
essential spectrum. To this end one can choose a smooth function $\chi(E)$ 
that vanishes outside of some interval contained in $(-mc^2,+mc^2)$ and 
then considers the Hamiltonian $\chi(\hat H_D)$. This now has a purely
discrete spectrum with eigenvalues $\chi(E_n)$, if the $E_n$'s are the 
eigenvalues of $\hat H_D$ with $|E_n|<mc^2$.

The same procedure can be applied to the other Hamiltonians, if situations
arise where their spectra are not purely discrete or when one is only
interested in certain spectral stretches. For the purpose of deriving a 
semiclassical trace formula one then considers a truncated time evolution
operator 
\begin{equation}
\label{timeevolvtrunc}
\hat U_\chi(t)=\ue^{-\frac{\ui}{\hbar}\hat H t}\,\chi(\hat H)\ ,
\end{equation}
whose spectral expansion in position representation reads
\begin{equation}
\label{Kernelexpan}
K_\chi(\vec{x},\vec{y},t) = \langle\vec{x}|\hat U_\chi(t)|\vec{y}\rangle 
=\sum_n \chi(E_n)\,\psi_n(\vec{x})\,\psi_n(\vec{y})^\dagger\,\ue^{-\frac{\ui}
{\hbar}E_n t}\ ,
\end{equation}
where $\psi_n(\vec{x})$ denotes the eigenspinor of $\hat H$ associated
with $E_n$.

The Gutzwiller trace formula has found many applications in the field of 
quantum chaos \cite{Gut90,QC} in which the principal questions are associated
with the distribution of eigenvalues and eigenfunctions of quantum Hamiltonians
in relation to properties of the corresponding classical dynamics. In this
context one first confines oneself to some spectral interval $I$ that 
contains $N_I<\infty$ eigenvalues; universal statistical properties then
emerge in the semiclassical limit $N_I\to\infty$. Here we implement this 
procedure by first choosing an interval
\begin{equation}
\label{Idef}
I = [E-\hbar\omega,E+\hbar\omega] \ ,
\end{equation}
and then performing the limit $\hbar\to 0$; and although the length of $I$ 
shrinks to zero, $N_I$ diverges in this limit (see below). In this approach
the localisation in energy described above hence appears to be very natural.

\subsection{Observables}
The quantum mechanical observables we consider are (bounded) Weyl operators,
i.e., they can be represented as in (\ref{HWeyl}). Their symbols $B(\vec{p},
\vec{x})$ are then functions on phase space taking values in the hermitian 
$n\times n$ matrices, and we suppose that they allow for asymptotic 
expansions in $\hbar$, 
\begin{equation}
\label{Basymptot}
B(\vec{p},\vec{x}) \sim \sum_{k\geq 0}\hbar^k\,B_k(\vec{p},\vec{x})\ .
\end{equation}
The precise meaning of such expansions is explained in \cite{Rob87}. The 
$\hbar$-in\-de\-pen\-dent term $B_0(\vec{p},\vec{x})$, the principal symbol
of the operator $\hat B$, represents the classical observable associated
with $\hat B$, at least concerning the translational degrees of freedom.
Expectation values of observables can be represented in terms of the 
symbol once one introduces a matrix valued Wigner transform of a spinor 
$\psi(\vec{x})=(\psi_1(\vec{x}),\dots,\psi_n(\vec{x}))^T$ through
\begin{equation}
\label{Wignerdef}
W[\psi]_{kl}(\vec{p},\vec{x}) := \int\ue^{-\frac{\ui}{\hbar}\vec{p}\cdot
\vec{y}}\overline{\psi_k}(\vec{x}-\tfrac{1}{2}\vec{y})\,\psi_l(\vec{x}
+\tfrac{1}{2}\vec{y})\ \ud y \ .
\end{equation}
Then
\begin{equation}
\label{expect}
\langle \psi,\hat B\psi \rangle = \frac{1}{(2\pi\hbar)^3}\int\int
{\rm tr}\left( W[\psi](\vec{p},\vec{x})\,B(\vec{p},\vec{x}) \right)\ 
\ud p\,\ud x\ . 
\end{equation}

\section{Trace Formula}
Gutzwiller's approach to the trace formula \cite{Gut71,Gut90} was to 
depart from a semiclassical expansion of the time evolution operator in 
position representation (\ref{Kernelexpan}). Expressing this in terms of a 
Feynman path integral and evaluating it in leading semiclassical order with 
the method of stationary phase, he arrived at a representation of the kernel 
$K(\vec{x},\vec{y},t)$ in terms of a sum over the classical trajectories 
connecting $\vec{y}$ and $\vec{x}$ in time $t$. In this context he made the 
important observation \cite{Gut67} that for not too small times $t$ each term 
in this sum must contain an extra phase factor that essentially consists of 
the Morse index of this trajectory.

Here we are interested in a trace formula for quantum systems with spin that 
yields a semiclassical expansion of a weighted and smeared spectral density, 
where the weights are provided by the expectation values of an observable 
$\hat B$ in the eigenstates of the Hamiltonian and the smearing ensures 
convergence of the sums involved, see also \cite{Bol00}. To this end the 
`spectral' side of the trace formula is given by 
\begin{equation}
\label{specTF}
\sum_n\chi(E_n)\, \langle \psi_n,\hat B\psi_n \rangle \, \rho\left(
\frac{E_n-E}{\hbar}\right) = {\rm Tr}\frac{1}{2\pi}\int_{-\infty}^{+\infty}
\tilde\rho(t)\,\ue^{\frac{\ui}{\hbar}Et}\,\hat B\,\hat U_\chi(t)\ \ud t\ .
\end{equation}
In this expression $\rho(E)$ is a smooth, regularising function with Fourier 
transform $\tilde\rho(t)$ that is also smooth and vanishes outside of some 
compact interval. The trace on the right-hand side will now be calculated
in position representation, after the leading semiclassical order of 
(\ref{Kernelexpan}), i.e., the appropriate Van Vleck-Gutzwiller propagator,
has been introduced.

\subsection{Van Vleck-Gutzwiller Propagator}
In an alternative approach to Gutzwiller's semiclassical expansion of
the propagator (\ref{Kernelexpan}), making use of a Feynman path integral, 
one represents the kernel in a way that is closely related to the WKB method. 
More precisely, one chooses the ansatz
\begin{equation}
\label{kernWKB}
K(\vec{x},\vec{y},t) = \frac{1}{(2\pi\hbar)^3}\int\sum_{k\geq 0}\left(
\frac{\hbar}{\ui}\right)^k\,a_k(\vec{x},\vec{y},t,\vecxi)\
\ue^{\frac{\ui}{\hbar}(S(\vec{x},\vecxi,t)-\vec{y}\cdot\vecxi)}\ \ud\xi\ ,
\end{equation}
and determines the phase $S(\vec{x},\vecxi,t)$ and the coefficients 
$a_k(\vec{x},\vec{y},t,\vecxi)$ of the amplitude by requiring that 
$K(\vec{x},\vec{y},t)$ solves the Schr\"odinger (Dirac, Pauli) equation to 
arbitrary powers in $\hbar$ with an appropriate initial condition at $t=0$. 
In the cases of the quantum Hamiltonians (\ref{HD}), (\ref{HP}) and 
(\ref{HPalt}) it turns out that the corresponding phases $S(\vec{x},\vecxi,t)$
have to be solutions of Hamilton-Jacobi equations with 
$H^\pm(\vec{p},\vec{x})$, $H_0(\vec{p},\vec{x})$ and $H^{\uparrow/\downarrow}
(\vec{p},\vec{x})$, respectively, as classical Hamiltonians (see 
(\ref{HDeigenv})--(\ref{HPalteigenv})). The matrix valued coefficients 
$a_k(\vec{x},\vec{y},t,\vecxi)$ of the amplitudes are determined by a 
hierarchy of (transport) equations that can be solved order by order in 
$\hbar$, starting with the leading expression $a_0(\vec{x},\vec{y},t,\vecxi)$,
which apart from the contribution of the translational motion contains the 
leading order of the spin dynamics. Knowing the solutions $S(\vec{x},\vecxi,t)$
and $a_0(\vec{x},\vec{y},t,\vecxi)$, one then still has to evaluate the 
integral (\ref{kernWKB}) over $\vecxi$ with the method of stationary phase. 
The details of this calculation can be found in \cite{BolKep99}.

Since $S(\vec{x},\vecxi,t)$ is a solution of a Hamilton-Jacobi equation, it 
is a generating function of a canonical transformation, $(\vec{p},\vec{x})
\mapsto (\vecxi,\vec{z})$. Here $\vec{p}=\nabla_x S(\vec{x},\vecxi,t)$ and
$\vec{z}=\nabla_\xi S(\vec{x},\vecxi,t)$, such that $(\vecxi,\vec{z})$ and
$(\vec{p},\vec{x})$ are starting and end points, respectively, of a solution
of the equations of motion generated by the respective classical Hamiltonian.
The stationary points $\vecxi_{\rm st}$ of the phase in (\ref{kernWKB})
are now uniquely related to classical trajectories from $\vec{y}$ to
$\vec{x}$, since the condition of stationarity reads $\vec{y}= \nabla_\xi 
S(\vec{x},\vecxi_{\rm st},t)$. The stationary point $\vecxi_{\rm st}$
itself hence is the momentum of the associated trajectory at time zero.
Thus the sum over the stationary points leads to a sum over classical
trajectories; see \cite{BolKep99} for the explicit form of these sums. In 
the cases of the Dirac- and Pauli-Hamiltonians (\ref{HD}) and (\ref{HP}) 
the trajectories that contribute are purely translational without 
contributions from spin (since the respective classical Hamiltonians 
contain no spin). The latter only contributes through weight factors
that come from the leading term $a_0(\vec{x},\vec{y},t,\vecxi_{\rm st})$ 
of the amplitude. Since this is evaluated at the stationary point 
$\vecxi_{\rm st}$, the spin contribution derives from a precession
along the associated classical trajectory. For a further discussion see
also the next subsection; more details are described in \cite{BolKep99}.

Only in the case of the Hamiltonian (\ref{HPalt}) is the spin already
directly contained on a classical level, see (\ref{HPalteigenv}). Here
one has to deal with the (two) translational dynamics of a particle whose 
spin is tight to the direction of the external `field' $\vec{D}$, such that
it follows this direction adiabatically along the trajectories of the 
particle. This adiabatic motion of the spin is reflected in the occurrence 
of a certain geometric phase, which has in the context of a Bohr-Sommerfeld 
quantisation been introduced by Littlejohn and Flynn \cite{LitFly91}, see 
also \cite{FriGuh93,BolKep99} for a discussion in the context of a trace 
formula.

\subsection{Semiclassical Spin Transport}
In the cases of the Dirac- and Pauli-Hamiltonians (\ref{HD}) and (\ref{HP})
there indeed is a dynamics of the spin degrees of freedom, which is driven
by the classical translational motion. These driven dynamics derive from 
the transport equation for the lowest order amplitude $a_0(\vec{x},\vec{y},t,
\vecxi_{\rm st})$ after separation of the purely translational part. The
spin transport equation that hence results reads \cite{BolKep99}
\begin{equation}
\label{spintranseq}
\dot{d}(\vec{p},\vec{x},t) + \ui\,\vec{C}(\vec{p}(t),\vec{x}(t))\cdot\vecsig\,
d(\vec{p},\vec{x},t) = 0 \ ,
\end{equation}
with initial condition $d(\vec{p},\vec{x},0)={\bf 1}_2$. Its solution, the 
spin transport matrix $d(\vec{p},\vec{x},t)\in{\rm SU}(2)$, propagates the 
(quantum) spin 1/2 along the classical trajectory $(\vec{p}(t),\vec{x}(t))$ 
starting at $(\vec{p},\vec{x})$. The vector $\vec{C}(\vec{p},\vec{x})$ 
depends on which Hamiltonian one considers and contains the fields $\vec{E}$ 
and $\vec{B}$; for a Pauli-Hamiltonian $\hat H_{\rm P}$ the quantity $\vec{C}$
is precisely the one described below (\ref{HP}). Moreover, if the vector 
$\vec{s}(t)$ denotes the expectation value of the (normalised) spin operator 
$\vecsig$ in a two-component spinor $u(t)=d(\vec{p},\vec{x},t)\,u(0)$, this 
`classical' spin obeys the equation
\begin{equation}
\label{spinprecess}
\dot{\vec{s}}(t) = \vec{C}(\vec{p}(t),\vec{x}(t)) \times \vec{s}(t)\ 
\end{equation}
of classical spin precession and thus provides an Ehrenfest relation for
the spin. In the relativistic case (\ref{spinprecess}) yields the well known 
Thomas precession, which Rubinow and Keller \cite{RubKel63} were the first to
derive semiclassically from the Dirac equation. 

One can now combine the Hamiltonian translational motion and the spin
dynamics that are driven by the former one into a single dynamical system
on a combined phase space, $(\vec{p}(0),\vec{x}(0),\vec{s}(0))\mapsto
(\vec{p}(t),\vec{x}(t),\vec{s}(t))$. In ergodic theory such combinations
are known as `skew products' of the two types of dynamics. In applications 
to quantum chaos the ergodic properties of precisely these combined
dynamics determine the `quantum chaotic' properties of the quantum system,
see \cite{BolKep99a,BolGla00,BolGlaKep01}.

\subsection{Semiclassical Trace Formula}
As mentioned earlier, the `semiclassical side' of the trace formula emerges
upon introducing the leading semiclassical order of the propagator 
(\ref{kernWKB}) on the right-hand side of (\ref{specTF}) and evaluating
all integrals, which involve the variables $(\vecxi,\vec{x},t)$, with the 
method of stationary phase. Since this method requires all stationary points 
$(\vecxi_{\rm st},\vec{x}_{\rm st},t_{\rm st})$ to be non-degenerate, the 
trace formula can only be derived under appropriate conditions on the 
classical systems. In particular, since the stationary points are such that
the phase space points $(\vecxi_{\rm st},\vec{x}_{\rm st})$ lie on a periodic 
orbit with energy $E$ (that appears on the left-hand side of (\ref{specTF})) 
and period $t_{\rm st}$, the conditions have indeed to be imposed on the 
periodic orbits. One such condition is that $E$ must not be a critical value 
of the relevant classical Hamiltonians, i.e., $(\nabla_p H(\vec{p},\vec{x}),
\nabla_x H(\vec{p},\vec{x}))\neq 0$ for all $(\vec{p},\vec{x})$ on the energy 
shell $\Omega_E=\{(\vec{p},\vec{x});\ H(\vec{p},\vec{x})=E\}$. This condition 
ensures that all stationary points with $t_{\rm st}=0$ are non-degenerate. The 
corresponding points $(\vecxi_{\rm st},\vec{x}_{\rm st})$ make up all of the 
energy shell $\Omega_E$ and the contribution of these stationary points yields
the leading semiclassical term (also called Weyl term) on the right-hand side 
of the trace formula. 

For the following we restrict our attention to the case where all non-trivial
periodic orbits $\gamma$ (i.e., $T_\gamma =t_{\rm st}\neq 0$) are isolated and 
non-degenerate. This means that their monodromy matrices $M_\gamma$, 
describing the linear stability of the orbits, have no eigenvalues one. This 
does not exclude elliptic orbits, if these are isolated, but only parabolic 
(marginally stable) ones. Furthermore, we give the trace formula explicitly 
for the case of a Dirac-Hamiltonian (\ref{HD}) and with the inclusion of an 
observable $\hat B$. The other cases follow from this trace formula by 
specialising to the appropriate simplified situations; e.g., the 
Pauli-Hamiltonian (\ref{HP}) has no contribution from anti-particles and 
therefore only contains one type of (translational) classical dynamics. In 
contrast, the Hamiltonian (\ref{HPalt}) does lead to two types of classical 
dynamics, generated by (\ref{HPalteigenv}), but has no independent spin 
dynamics; there only is an additional geometric phase as described in 
\cite{LitFly91,FriGuh93,BolKep99}. Now the trace formula reads
\cite{BolKep99,Bol00}
\begin{eqnarray}
\label{DiracTF}
\sum_n\chi(E_n)\,& &\hspace*{-4mm}\langle \psi_n,\hat B\psi_n \rangle \, 
\rho\left(\frac{E_n-E}{\hbar}\right) \\
  &=& \frac{\tilde\rho(0)}{2\pi}\frac{\chi(E)}{(2\pi\hbar)^2}\left(2\,{\rm vol}
      \Omega_E^+\,{\rm tr}\overline{B_0^+}^E + 2\,{\rm vol}\Omega_E^-\,{\rm tr}
      \overline{B_0^-}^E \right) +O(\hbar^{-1}) \nonumber \\
  & & +\sum_{\gamma^{\pm}}\chi(E)\,\frac{\tilde\rho(T_{\gamma^\pm})}
      {2\pi}\,{\rm tr}\overline{B_0}^{\gamma^{\pm}}\,A_{\gamma^\pm}\,
      \ue^{\frac{\ui}{\hbar}S_{\gamma^\pm}(E)}\,(1+O(\hbar)) \ . \nonumber
\end{eqnarray}
The first term on the right-hand side is the so-called Weyl term and yields
the leading semiclassical approximation of  the left-hand side. It essentially
contains the averages $\overline{B_0^\pm}^E$ of the principal symbol 
$B_0(\vec{p},\vec{x})$ of the observable over the two energy shells 
$\Omega_E^\pm$. The two contributions are weighted according to the relative 
volumes of the respective energy shells. The characteristic quantities 
appearing in the sum over the two types $\gamma^\pm$ of periodic orbits are 
the same as in the case of the Gutzwiller trace formula for the spectral 
density of a quantum system without spin: there is an exponential factor with 
the actions $S_{\gamma}(E)$, and an amplitude that contains the well known
part
\begin{equation}
\label{Gutzamp}
A_{\gamma} = \frac{T_{\gamma}^{\rm prim}\,\ue^{-\ui\frac{\pi}{2}
\mu_{\gamma}}}{|\det(M_\gamma -{\bf 1})|^{1/2}} \ ,
\end{equation}
with the associated primitive period, the Maslov index $\mu_{\gamma}$, and
the stability denominator. The additional factors come from the regularisation
and from the presence of the observable. In particular, 
$\overline{B_0}^{\gamma^{\pm}}$ denotes an average of the projection of
$B_0(\vec{p},\vec{x})$ to one of the eigenspaces corresponding to the 
eigenvalues (\ref{HDeigenv}), weighted with the spin transport matrix 
$d(\vec{p},\vec{x},t)$, along the periodic orbit $\gamma^{\pm}$. Details of 
this average are described in \cite{Bol00}.

A trace formula for a truncated spectral density, with only contributions
from the discrete spectrum of $\hat H_{\rm D}$, can be obtained from 
(\ref{DiracTF}) by choosing the observable $\hat B={\rm id}$ and removing 
the regularisation provided by the function $\rho$. The result is
\cite{BolKep99}
\begin{eqnarray}
\label{TFdense}
d_\chi(E) &=& \sum_n \chi(E_n)\,\delta(E_n -E) \nonumber \\
   &=& \chi(E)\,\frac{2\,{\rm vol}\Omega_E^+ +2\,{\rm vol}\Omega_E^-}
       {(2\pi\hbar)^3} + O(\hbar^{-2}) \\
   & & +\frac{\chi(E)}{2\pi\hbar}\sum_{\gamma^\pm}\frac{T_{\gamma^\pm}^{\rm 
       prim}\,{\rm tr}d_{\gamma^\pm}}{|\det(M_{\gamma^\pm}-{\bf 1})|^{1/2}}\,
       \ue^{\frac{\ui}{\hbar}S_{\gamma^\pm}(E)-\ui\frac{\pi}{2}
       \mu_{\gamma^\pm}}\,(1+O(\hbar)) \ , \nonumber
\end{eqnarray}
where here $d_{\gamma^\pm}$ denotes the spin transport matrix associated
with the transport of spin once along the periodic orbit. One immediately
observes that in contrast to the Gutzwiller trace formula without spin
two types of classical dynamics (particles and anti-particles) occur, and
spin contributes a factor of two in the Weyl term as well as weights
${\rm tr}\,d_{\gamma^\pm}$ for the periodic orbits describing the effect of 
spin transport. 

Obviously, the Weyl term in (\ref{TFdense}) yields the 
semiclassically leading contribution to the spectral density. This allows 
to derive the leading semiclassical asymptotics of the number $N_I$ of 
eigenvalues in the interval (\ref{Idef}), which is given by $2\hbar\omega$ 
times the Weyl term, and hence $N_I$ diverges as $\hbar^{-2}$ in the 
semiclassical limit.

\section{Applications}
Many applications of the Gutzwiller trace formula originate from problems
in the field of quantum chaos (see, e.g., \cite{Gut90,QC}). Apart from the
question for semiclassical quantisation rules (see, e.g., \cite{focus})
one of the major successes in this field was Berry's semiclassical
analysis of spectral two-point correlations \cite{Ber85} based on the
trace formula. In a certain range of validity, which stems from the so-called
diagonal approximation Berry employed, he verified that the two-point
correlations of energy levels of classically chaotic quantum systems (without 
spin) follow the predictions of random matrix theory (RMT). Subsequently, 
Bogomolny and Keating \cite{BogKea96} extended Berry's result in that they 
went one step beyond the diagonal approximation. 

The first application of trace formula techniques, without, however, having 
a complete trace formula available, to quantum systems with spin goes back 
to Frisk and Guhr \cite{FriGuh93}. They considered a quantum Hamiltonian of
the type (\ref{HPalt}) describing spin-orbit coupling in certain billiards.
In spirit, they applied the trace formula in `reverse direction' in that
they used quantum energy levels in order to obtain information about the
contribution of various types of periodic orbits. Similar studies can be
found in \cite{BraAma00}.   

A second type of applications concerns an extension of the semiclassical 
analysis of spectral two-point correlations to quantum systems with spin 1/2.
If a time-reversal symmetry is present, Kramers' degeneracy implies that
all energy levels of a quantum system with half-integer spin are doubly
degenerate. After removal of this systematic multiplicity, the spectral
statistics should be described by the Gaussian symplectic ensemble (GSE)
of RMT, if the corresponding classical system is chaotic. Without 
time-reversal symmetry Kramers' degeneracy is absent, and the relevant
ensemble of RMT is the Gaussian unitary one (GUE). In \cite{BolKep99a}
Berry's semiclassical approach is carried out with the trace formula
(\ref{DiracTF}), or (\ref{TFdense}), as a basis. It is shown that, within
the same range of validity as in \cite{Ber85}, indeed the two-point 
correlations agree with the GSE or GUE, respectively. Moreover, in
\cite{Kep00} the Bogomolny-Keating method is carried over to the case
of spin 1/2, with the same findings. In both studies, apart from a chaotic
translational motion, ergodic properties of the combined dynamics described
below (\ref{spinprecess}) are needed in order to obtain an agreement with the 
RMT predictions.

The Weyl term of the trace formula (\ref{DiracTF}) moreover allows to
determine a semiclassical average of the expectation values $\langle\psi_n,
\hat B\psi_n\rangle$ in eigenstates of $\hat H_{\rm D}$ with $E_n\in I$. The
result is given by \cite{Bol00}
\begin{equation}
\label{Szegoe}
\lim_{\hbar\to 0}\frac{1}{N_I}\sum_{E_n\in I}\langle\psi_n,\hat B\psi_n\rangle
=\frac{1}{2}\frac{{\rm vol}\Omega_E^+\,{\rm tr}\overline{B_0^+}^E + {\rm vol}
\Omega_E^-\,{\rm tr}\overline{B_0^-}^E}{{\rm vol}\Omega_E^+ + {\rm vol}
\Omega_E^-} \ ,
\end{equation}
and does not require the classical system to be chaotic. The only requirement
is that the set of periodic orbits on both energy shells must be of
measure zero, which is a comparatively weak condition. Only if one requires
individual expectation values to approach the expression on the right-hand side
of (\ref{Szegoe}) as $\hbar\to 0$ one needs a stronger condition on the
classical side, since then no cancelations on the left-hand side are allowed.
For the case of Pauli-Hamiltonians (\ref{HP}) it has been proven 
\cite{BolGla00} (see also \cite{BolGlaKep01}) that almost all expectation 
values indeed converge to the equivalent of the right-hand side of 
(\ref{Szegoe}), if the combined dynamics of translational and spin degrees 
of freedom are ergodic. For quantum systems without spin such a result had 
been known before under the notion of quantum ergodicity (see, e.g., 
\cite{HelMarRob87}).
%
%
%

%
\end{document}